\documentclass[aps,prb,twocolumn,floatfix]{revtex4}
\usepackage{amssymb}
\usepackage{graphicx,subfigure}
\usepackage{bm}

\begin{document}

\title{Properties of a diagonal 2-orbital ladder model of the Fe-pnictide superconductors}
\date{\today }
\author{ E. Berg$^{1,2}$, S. A. Kivelson$^{2}$, and D.~J.~Scalapino$^{3}$}
\affiliation{$^1$Department of Physics, Harvard University, Cambridge, MA 02138, USA\\
$^2$Department of Physics, Stanford University, Stanford, CA 94305-4045, USA%
\\
$^3$Department of Physics, University of California, Santa Barbara, CA
93106-9530, USA}

\begin{abstract}
We study a diagonal 2-orbital ladder model of the Fe based
superconductors using the density matrix renormalization group
method. At half filling, we find a close competition between a
``spin-striped'' state and a non-collinear ``spin-checkerboard''
state, as well as significant nematic correlations. Upon finite
hole or electron doping, the dominant pairing correlations are
found to have A$_{1,g}$ ($S-$wave) symmetry.
\end{abstract}

\maketitle

The recent discovery
of iron pnictide superconductors
\cite{Kamihara-short} has added to the list of materials for which
the superconducting pairing mechanism appears to be of electronic
origin.
From a theoretical point of view, these materials provide us with a rare
opportunity to test our understanding of unconventional superconducting
mechanisms. Numerous models have been proposed for these materials. In order
for these models to be 
more tractable, most authors have taken either a weak\cite
{Mazin2-short,Kuroki,Raghu-short,Wang-short,Chubukov,Graser-short,tesanovic}
or a strong \cite{Fang-short,Xu,Hu,Kruger} coupling starting point
(\emph{i.e}. assuming that the interaction strength is either much
smaller or much larger than
the bandwidth). However, there is evidence that the
actual materials are in the \emph{ intermediate coupling}
regime\cite{Si1,Si2-short,Haule2008,Aichhorn-short,Qazilbash-short},
which is
also the most difficult to treat 
analytically. In this regime, one has to resort to numerical methods\cite%
{Daghofer-short,Haule2008,Aichhorn-short,Moreo-short,Berg}, which
are currently limited to either small clusters or to one
dimensional systems. Despite these limitations, one may still hope
that the important ordering tendencies of the system (which are
presumably driven by short
range microscopic energetics) may 
be apparent already for small size systems.

In this Paper, we study a 
2-orbital diagonal ladder model
of the pnictides. The model is
solved using the density matrix renormalization group
(DMRG)\cite{white} technique, which enables us to study the
sensitivity of the results to the system size in one direction.
The diagonal geometry has the advantage that it preserves the
symmetry between the $x$ and $y$ directions, thus enabling us to
address some of the outstanding questions of the field, such as
the question of the gap symmetry and of nematic ordering. For
example, the reflection
symmetry of the model makes the distinction between A$_{1g}$--like ($S-$%
wave) and B$_{2g}$--like ($D_{x^{2}-y^{2}}-$wave) precise.
\begin{figure}[t]
\includegraphics[width=0.5\textwidth]{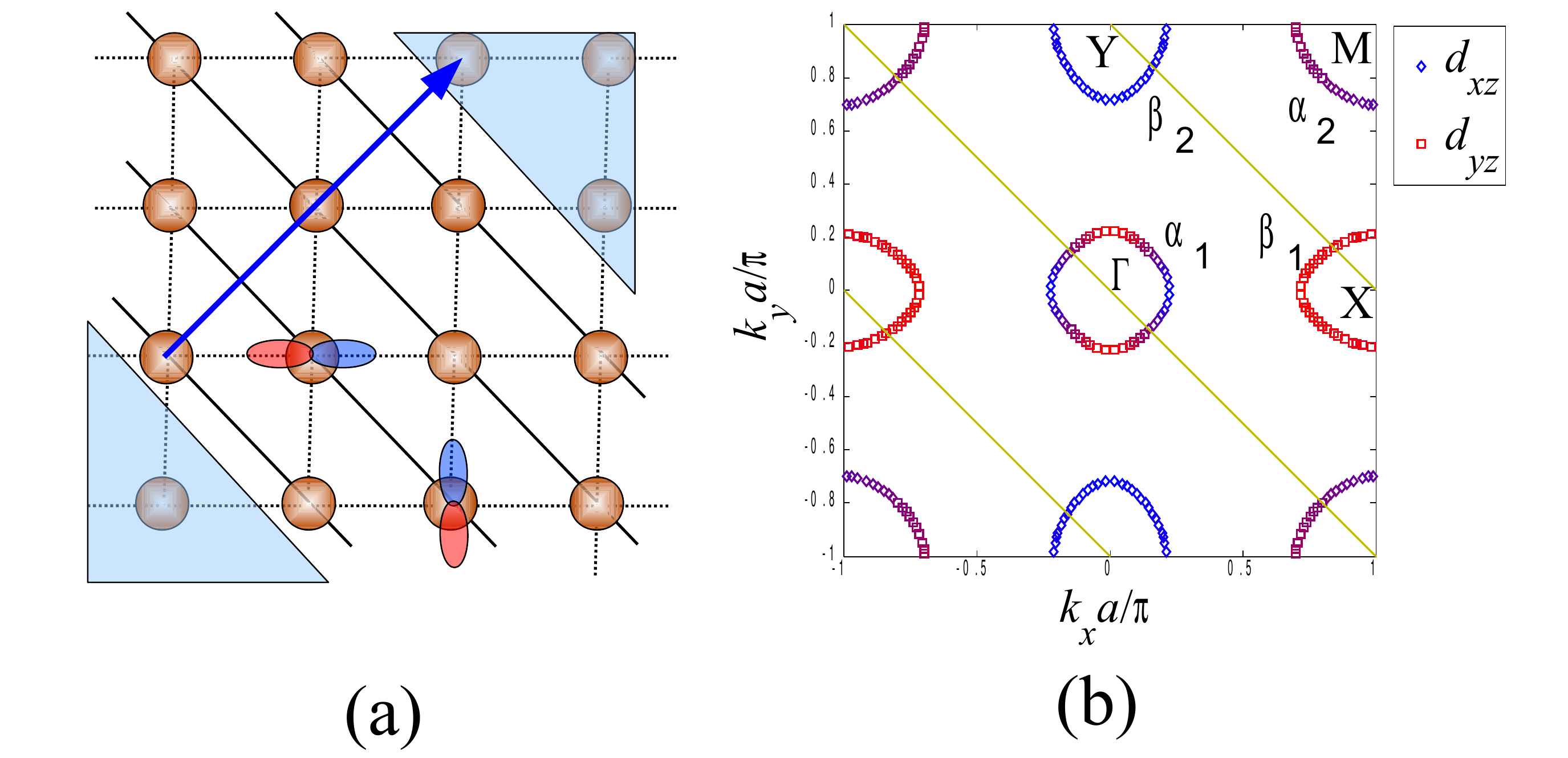}
\caption{(Color online.) (a) The diagonal ladder geometry. Each
site has two orbitals, $d_{xz}$ and $d_{yz}$. Sites separated by
$(2a,2a)$ (indicated by the arrow in the figure) are identified.
(b) the Brillouin zone of the two-band
model\protect\cite{Raghu-short} in the unfolded (one Fe/unit cell)
scheme. The Fermi surfaces are shown, along with their orbital
content. The diagonal lines show the k-states which can be
accessed in the diagonal ladder.} \label{fig:1}
\end{figure}

\emph{Model--} The geometry of the model is shown in Fig. \ref{fig:1}.
Starting from the 2D square Fe lattice, we cut out four parallel chains
directed along $(1,-1)$. We then impose periodic boundary conditions in the
transverse direction, such that Fe atoms separated by $\Delta \mathbf{R=}%
2a\left( 1,1\right) $ are identified. ($a$ is the Fe-Fe spacing.) For each
Fe site we keep one $d_{xz}$ and one $d_{yz}$ orbital. The Hamiltonian
is written as $H=H_{0}+H_{int}$, where\cite{Raghu-short} 
\begin{eqnarray}
H_{0} &=&\sum_{\mathbf{R},\sigma }{\Big[}-t_{1}d_{x\sigma ,\mathbf{R}%
}^{\dagger }d_{x\sigma ,\mathbf{R}+a\mathbf{\hat{x}}}^{\vphantom{\dagger}%
}-t_{2}d_{x\sigma ,\mathbf{R}}^{\dagger }d_{x\sigma ,\mathbf{R}+a\mathbf{%
\hat{y}}}^{\vphantom{\dagger}}  \nonumber \\
&&+\sum_{\xi =\pm 1}\left( -t_{3}d_{x\sigma ,\mathbf{R}}^{\dagger
}d_{x\sigma ,\mathbf{R}+\xi \mathbf{\eta }}^{\vphantom{\dagger}%
}-\xi t_{4}d_{x\sigma ,\mathbf{R}}^{\dagger }d_{y\sigma ,\mathbf{R}+\xi \mathbf{%
\eta }}^{\vphantom{\dagger}}\right)   \nonumber \\
&&+\mathrm{H.c.}-\mu n_{x,\mathbf{R}}+\left( x\leftrightarrow y\right) {\Big]%
}\text{,}  \label{H0}
\end{eqnarray}%
where $\mathbf{\eta }=\left( a,a\right) $, and
\begin{eqnarray}
&&H_{\mathrm{int}}=\sum_{\mathbf{R}}{\Big[}\sum_{\alpha =x,y}Un_{\alpha
\uparrow ,\mathbf{R}}n_{\alpha \downarrow ,\mathbf{R}}+Vn_{x\mathbf{R}}n_{y%
\mathbf{R}}  \label{Hint} \\
&-&J\mathbf{S}_{x,\mathbf{R}}\cdot \mathbf{S}_{y,\mathbf{R}}+J^{\prime
}\left( d_{x\mathbf{\uparrow },\mathbf{R}}^{\dagger }d_{x\mathbf{\downarrow }%
,\mathbf{R}}^{\dagger }d_{y\mathbf{\downarrow },\mathbf{R}}^{%
\vphantom{\dagger}}d_{y\mathbf{\uparrow },\mathbf{R}}^{\vphantom{\dagger}}+%
\mathrm{H.c.}\right) {\Big]}\text{.}  \nonumber
\end{eqnarray}%
Here, $d_{\alpha \sigma ,\mathbf{R}}^{\dagger }$ with $\alpha =x,y$
creates an electron at site $\mathbf{R}$ with spin $\sigma $ in the $d_{xz}$ and $%
d_{yz}$ orbital, respectively. We have defined $n_{\alpha \sigma
}=d_{\alpha \sigma }^{\dagger }d_{\alpha \sigma }^{\vphantom{\dagger}}$, $%
n_{\alpha }=n_{\alpha \uparrow }+n_{\alpha \downarrow }$ and $\mathbf{S}%
_{\alpha }=\sum_{\sigma \sigma ^{\prime }}d_{\alpha \sigma }^{\dagger }\frac{%
\mathbf{\tau }_{\sigma \sigma ^{\prime }}}{2}d_{\alpha \sigma ^{\prime }}^{%
\vphantom{\dagger}}$ where
$\mathbf{\tau }$ are Pauli matrices. In Eq. (\ref{H0}) we have
used the tight binding parameters
which were obtained in Ref. \cite{Raghu-short}: $t_{1}=-1$, $t_{2}=1.3,$ $%
t_{3}=t_{4}=-0.85$. (We will henceforth measure all energies in units of $%
\left\vert t_{1}\right\vert $.) The interaction parameters in Eq. (\ref{Hint}%
) were chosen to satisfy the constraints $U-V=\frac{5}{4}J$,
$J^{\prime }=J/2$ which follow from assuming a rotationally
invariant Fe atom. (To check the sensitivity of the results to
parameters, we repeated the calculations using $U-V=J$. The
results did not change qualitatively.) In order to reduce the
number of parameters, we fixed $J=\frac{U}{4}$. Most of the
calculations described below were done with $U=4
- 8$, which is in the
\textquotedblleft intermediate coupling\textquotedblright\ regime
(such that $U$ is smaller than the overall bandwidth, but larger
than the Fermi energy of the electron and hole pockets).

In k-space, the diagonal ladder geometry can be thought of as cutting
through the 2D (one Fe/unit cell) Brillouin zone 
along the lines $\mathbf{k}=k_{1}\left( 1,-1\right) +k_{2}\left( 0,1\right) $%
, where $k_{1}\in \left[ -\frac{\pi }{a},\frac{\pi }{a}\right] $
and $k_{2}=0 $ or $\frac{\pi }{a}$. The resulting allowed points
in k-space are shown in Fig. \ref{fig:1}b, along with the Fermi
surface of the 2-band model at half filling (one electron per
orbital). The 2-orbital model has several well--known
shortcomings:  the $\alpha _{2}$ pocket is centered at $\left(
\frac{\pi }{a},\frac{\pi }{a}\right)$, while density functional
theory calculations show 
it at $\left( 0,0\right)$, and the $d_{xy}$ contribution which appears on
parts of the Fermi surface is missing. However, as we will discuss, the
important interplay between the $d_{xz}$ and $d_{yz}$ is taken into account.
\begin{figure}[t]
\includegraphics[width=0.5\textwidth]{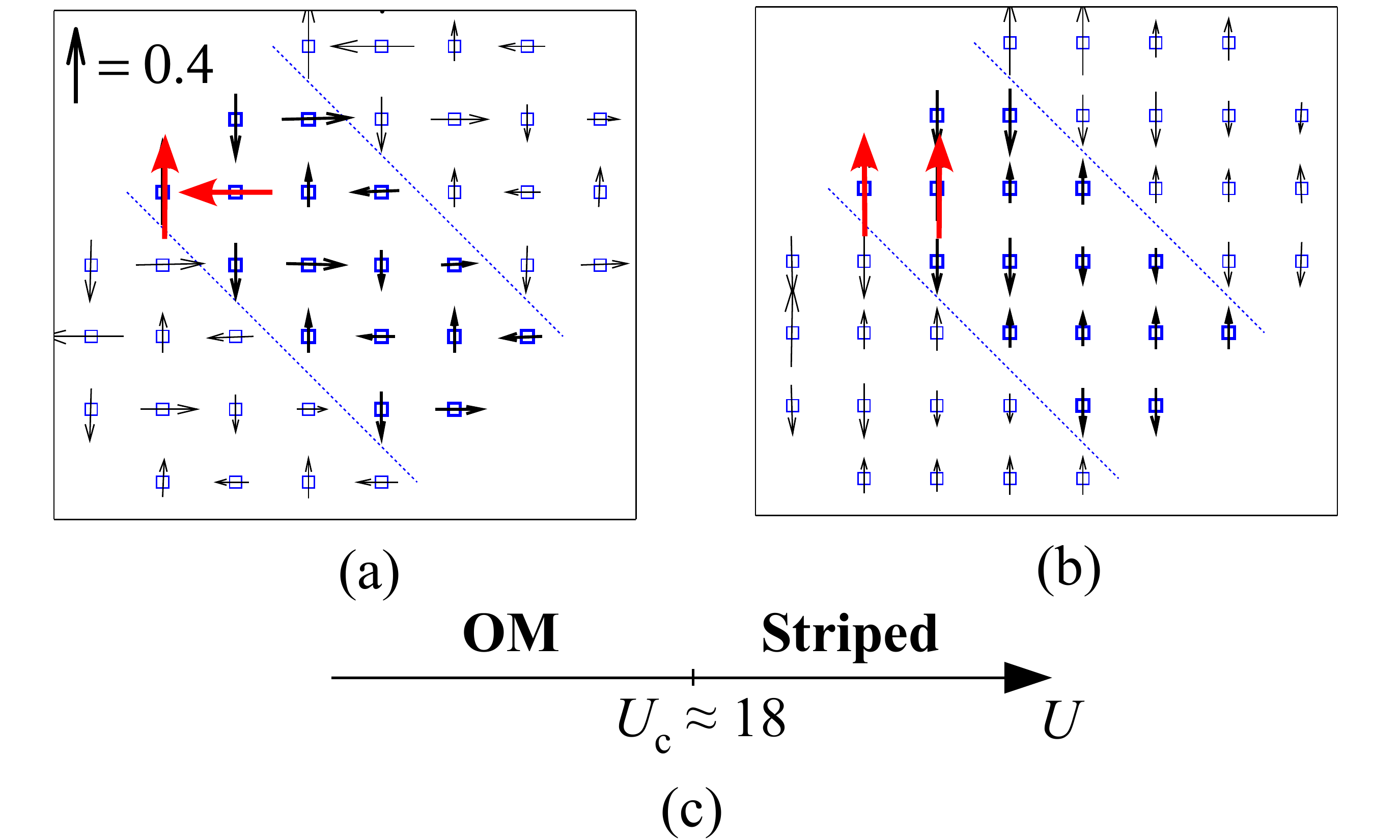}
\caption{(Color online.) (a,b) Spin patterns in $4\protect\sqrt{2}a\times
\protect\sqrt{2}a$ half filled (two electrons per site) systems. The arrows
are proportional to the local $\langle \mathbf{S}\rangle$. The diagonal
dashed lines show the actual boundaries of the system. Zeeman fields of
strength $h=0.5$ where applied on the $(0,0)$ and $(1,0)$ sites, in the
directions indicated by the bold arrows. (c) A schematic phase diagram at
half filling.}
\label{fig:2}
\end{figure}

The diagonal ladder has 
translational symmetry 
and reflection symmetry with respect to two mirror planes which are formed
by the $z$ axis and
a line in the $(1,\pm 1)$ directions that passes through
a site. (Note that these reflection operations interchange the $d_{xz}$
and $%
d_{yz}$ orbitals.) Therefore, the diagonal ladder
can support a \textquotedblleft nematic\textquotedblright\ phase in which
the symmetry between the $x$ and $y$ directions is spontaneously broken. In
addition, 
there is a sharp symmetry distinction between an A$_{1g}$ 
and a B$_{1g}$ 
superconducting order parameter: A$_{1g}$ 
is even 
under reflection 
through the mirror planes, and B$_{1g}$ is odd. A$_{1g}$ and
B$_{2g}$ ($D_{xy}$) are not distinct, since they are both even
under reflection. However, one can still distinguish between
A$_{1g}$-- and B$_{2g}$--like order parameters according to the
relative sign of the order parameter on $(1,1)$ and $(1,-1)$
oriented bonds.

\emph{Magnetic and nematic correlations}-- We begin from the half filled
case (one electron per orbital). In order to study the ordering tendencies
of the system, we apply various types of symmetry breaking perturbations at
the edge and study how they propagate into the bulk.

In the DMRG\ calculations described below, we have kept up to $m=6000$
states in situations where both the number of particles and the $z$
component of the total spin are conserved, and up to $3600$ states in cases
where one of these conservation laws is not present.
The maximum truncation error was less than $4\cdot 10^{-4}$ in all cases.%

Fig. \ref{fig:2} shows the expectation value of the total magnetization $%
\mathbf{m}=\sum_{\alpha }\langle \mathbf{S}_{\alpha }\rangle $ as a function
of position in 
a $4\sqrt{2}a\times \sqrt{2}a$ system 
with $U=8$. The total number of sites is $16$ ($32$ orbitals). In these
calculations, a Zeeman field term of the form $-\mathbf{h}_{\mathbf{R}}%
\mathbf{\cdot S}_{\mathbf{R}} $ was applied to two sites near the upper left
edge, $\mathbf{R=}\left( 0,0\right) $ and $\left( a,0\right) $. In Fig. \ref%
{fig:2}a, the fields were 
$\mathbf{h}_{\left( 0,0\right) }=h\mathbf{\hat{z}}$ and $\mathbf{h}_{\left(
a,0\right) }=-h\mathbf{\hat{x}}$,
while in Fig. \ref{fig:2}b the fields were
$\mathbf{h}_{\left( 0,0\right) }=\mathbf{h}_{\left( a,0\right) }=h\mathbf{%
\hat{z}}$. The magnitude of the fields was $h=0.5$.
As can be seen in the figure, these edge fields pin very different ordering
patterns. The pattern in Fig. \ref{fig:2}b 
is the \textquotedblleft spin striped\textquotedblright\ pattern with
momentum $\left( 0,\frac{\pi }{a}\right) $, while in \ref{fig:2}a 
we find a \textquotedblleft checkerboard\textquotedblright\ phase in which
the magnetization
on each of the two sublattices
is orthogonal to 
the other, and 
both $\left( \frac{\pi }{a},0\right) $ and $\left( 0,\frac{\pi }{a}\right) $
momentum components are present. This phase has been found in unrestricted
Hartree-Fock calculations \cite{Lorenzana-short}, and has been termed
\textquotedblleft orthomagnetic\textquotedblright\ (OM). The ground state
energies in these two 
ordered states are
equal within our numerical accuracy. In order to determine which of these
states is the true groundstate, we have undertaken two independent
approaches. We have computed the ground-state energy of systems with $N=8- 16
$ sites, in the presence of a bulk ordering field of strength $h=0.5$ on
each site, with a pattern of orientations 
which forces the spin-orders in Figs. \ref{fig:2}(a,b). For $U
\lesssim 18$, we have found for all system sizes studied that the
energy of the OM pattern
is lower than that of the striped pattern by a small amount of about $%
10^{-3} $ per site. Conversely, for larger $U$ we find that the
stripe pattern has a lower energy. We have obtained similar
results from an analysis of the ``spin-nematic'' order,
$\mathcal{N}_{\mathbf{R}} =\langle
3(S^z_{\mathbf{R}})^2 - \mathbf{S}^2_{\mathbf{R}}\rangle$ measured on sites $%
\mathbf{R}$ on one sublattice with a staggered Zeeman field, $-hS^z$ applied
to the other sublattice; positive values of $\mathcal{N}_{\mathbf{R}}$ are
indicative of stripe and negative values of OM ordering tendencies.
\begin{figure}[t]
\includegraphics[width=0.51\textwidth]{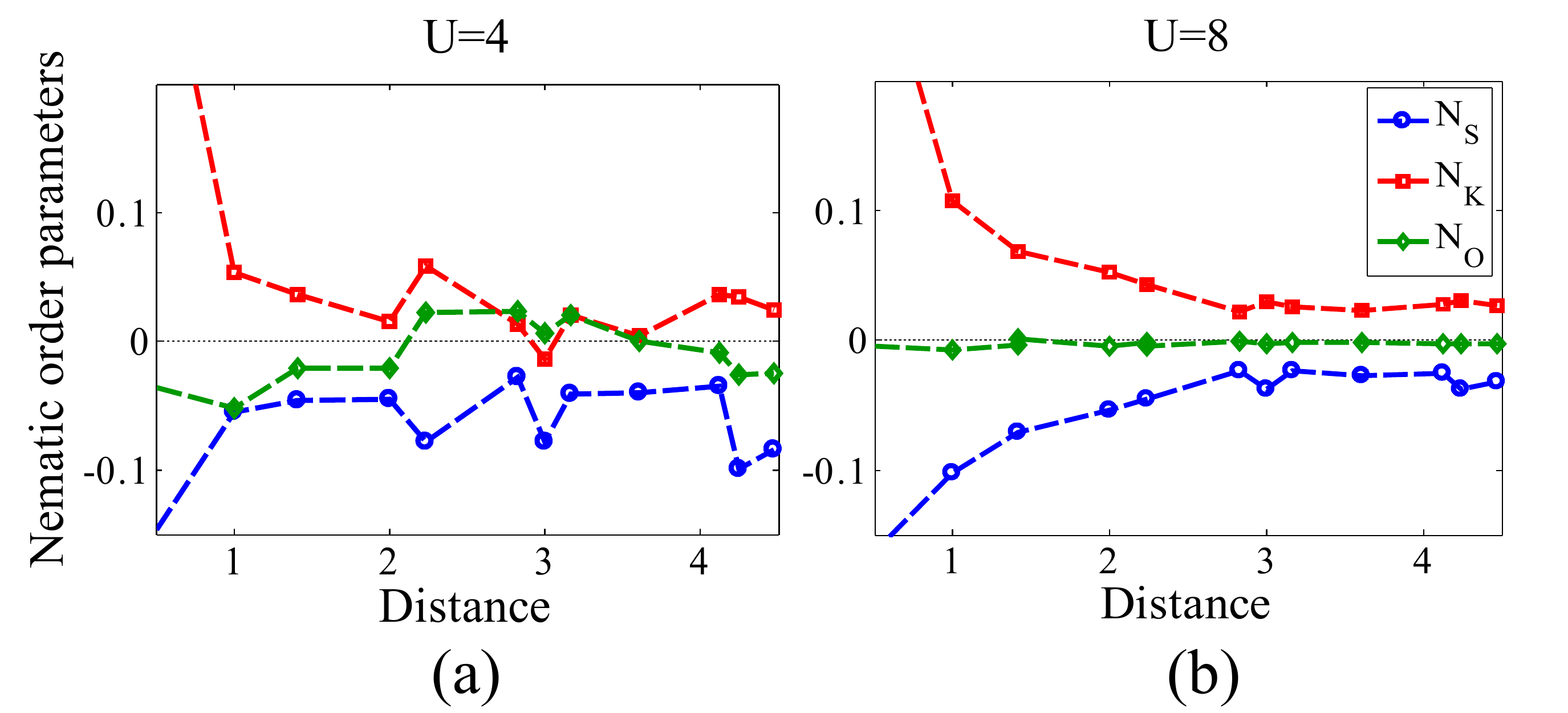}
\caption{(Color online.) Measurements of the nematic order parameters [see
Eq. (\protect\ref{eq:nematic})] as a function of distance from the origin,
in $4\protect\sqrt{2}a\times \protect\sqrt{2}a$ (16 site) systems. (The
geometry is the same as in the clusters shown in Fig. \protect\ref{fig:2}.)
(a) $U=4$, (b) $U=8$. In these calculations, the hopping strengths of the
bond from $(0,0)$ to $(a,0)$ (at the upper left corner of the system) were $%
50\%$ stronger than in the bulk.}
\label{fig:3}
\end{figure}

The fact that the OM\ and the striped states are nearly degenerate
can be understood as a consequence of the magnetic frustration of
the exchange interactions between the two
sublattices\cite{Si1,Fang-short,Xu}. In the strong coupling limit,
$U \gg 1$, the 2-band model maps onto a spin-1 Heisenberg model
with a nearest-neighbor exchange interaction $J_{1}$ and a
next-nearest neighbor $J_{2}$. In the regime $J_{2}>0.5J_{1}$, the
classical ($S\to \infty$) ground state consists of
antiferromagnetically aligned $A$ and $B$ sublattices, while their
relative orientation is completely free.
However, $%
1/S$ quantum fluctuations favor the striped state over the OM state.
This accounts for the weak preference for stripe order for $U
\gtrsim 18$. However,  for somewhat smaller $U$, biquadratic spin
interaction terms of
the form $K\left( \mathbf{R}_{1},\dots ,\mathbf{R}_{4}\right) \left( \mathbf{%
S}_{\mathbf{R}_{1}}\cdot \mathbf{S}_{\mathbf{R}_{2}}\right) \left( \mathbf{S}%
_{\mathbf{R}_{3}}\cdot \mathbf{S}_{\mathbf{R}_{4}}\right) $, are generated.
Since their magnitude is of order $\frac{t^{4}}{U^{3}}\sim J^2/U$, they
are negligible in the 
large--$U$ limit, but (at least for the parameters we have
explored) they produce a weak preference for the OM state for $U <
18$.

In the parent FeAs compounds, the ground state has a striped spin pattern,
in contrast to our model in the realistic intermediate-coupling regime.
Since the OM\ and the striped states are extremely close in energy, it
is easy
to understand how small perturbations, \emph{e.g.} slightly different model
parameters or the coupling to the lattice, can stabilize the striped state
relative to the OM\ state.

\begin{figure}[t]
\includegraphics[width=0.5\textwidth]{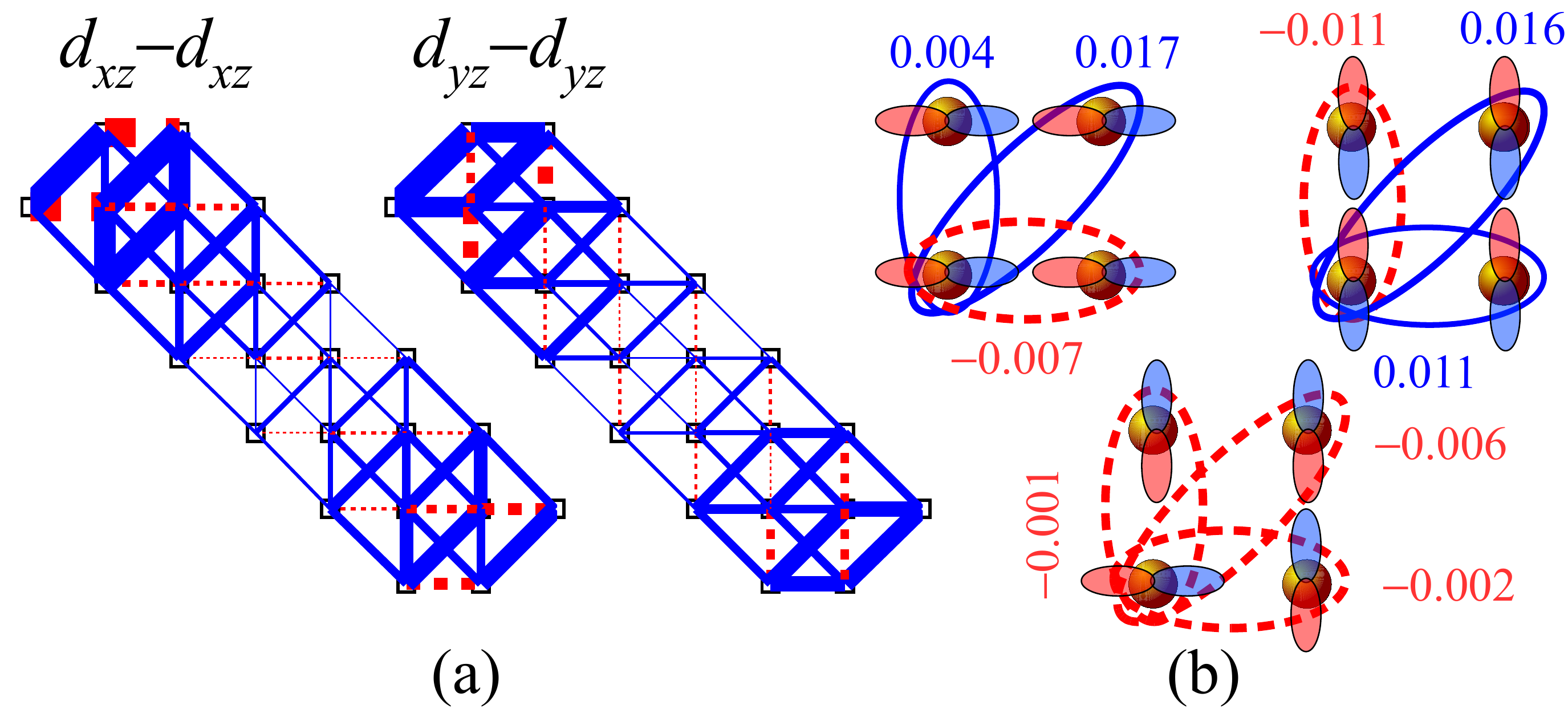}
\caption{(Color online.) (a) Bond pairing amplitudes for a $6\protect\sqrt{2}%
a\times \protect\sqrt{2}a$ (24 site) system with $U=4$, doped with
6 holes. Left: $d_{xz}-d_{xz}$ pairing, right: $d_{yz}-d_{yz}$
pairing. An external pairing field of strength $\Delta=0.5$ is
applied to the upper left diagonal $d_{xy}-d_{xy}$ bond [see Eq.
(\protect\ref{eq:H1})], highlighted in the figure. The thickness
of the bonds represents the amplitude of induced pair field on
that bond. Solid (dashed) lines stand for positive (negative)
amplitudes, respectively. (b) Detailed pair structure near the
middle of the system. Upper left: $d_{xz}-d_{xz}$ amplitudes,
upper right: $d_{yz}-d_{yz}$, and bottom: $d_{xz}-d_{yz}$
amplitudes.} \label{fig:4}
\end{figure}

The striped state breaks the $C_{4}$ symmetry of the square 2D lattice down
to $C_{2}$, and thus this state has a \textquotedblleft
nematic\textquotedblright\ component. In the diagonal ladder geometry, the
nematic character appears as a breaking of reflection symmetry about the $%
\left( 1,\pm 1\right) $ directions. It is therefore interesting to calculate
the nematic response of the system. We define the nematic order parameters:
\begin{eqnarray}
N_{\mathrm{S}}\left( \mathbf{R}\right) &=&\sum_{\alpha }\langle \lbrack
\mathbf{S}_{\alpha ,\mathbf{R}}\cdot \mathbf{S}_{\alpha ,\mathbf{R}+a\mathbf{%
\hat{x}}}-\mathbf{S}_{\alpha ,\mathbf{R}}\cdot \mathbf{S}_{\alpha ,\mathbf{R}%
-a\mathbf{\hat{y}}}]\rangle \text{,}  \nonumber \\
N_{\mathrm{K}}\left( \mathbf{R}\right) &=&2\sum_{\alpha ,\sigma
}\Big{[}|\langle
d_{\alpha ,\sigma ,\mathbf{R}}^{\dagger }d_{\alpha ,\sigma ,\mathbf{R}+a%
\mathbf{\hat{x}}}^{\vphantom{\dagger}}\rangle |-|\langle d_{\alpha ,\sigma ,%
\mathbf{R}}^{\dagger }d_{\alpha ,\sigma ,\mathbf{R}-a\mathbf{\hat{y}}}^{%
\vphantom{\dagger}}\rangle |\Big{]}  \nonumber \\
N_{\mathrm{O}}\left( \mathbf{R}\right) &=&\langle \lbrack n_{x,\mathbf{R}%
}-n_{y,\mathbf{R}}]\rangle \text{.}  \label{eq:nematic}
\end{eqnarray}%
These order parameters were measured in a calculation in which the hopping
strength on the bond from $\left( 0,0\right) $ to $\left( a,0\right) $ was
enhanced by $50\%$ relative to the bulk, thus breaking reflection symmetry
about $(1,-1)$ locally.
Fig. \ref{fig:3} shows the order parameters as a function of the distance
from the 
boundary at which the perturbation is applied.

After an initial decay, $N_S$ and $N_K$ saturate and remain nearly constant.
For $U=4$, there is some degree of ``orbital order'' $N_O$, which seems to
fluctuate around zero. For $U=8$, $N_O$ is very small, as can be expected
from the fact that strong repulsive interactions suppress both intra-- and
inter--orbital density fluctuations. The substantial nematic correlations
reflect the closeness in energy of the spin-striped state
to the ground state.


\emph{Pairing correlations--} In order to study the pairing
response of the system, we have added a boundary pairing potential
of the form
\begin{equation}
H_{1}=-\Delta \left[ d_{x,\uparrow ,\mathbf{R}_{1}}^{\dagger
}d_{x,\downarrow ,\mathbf{R}_{2}}^{\dagger }-d_{x,\downarrow ,\mathbf{R}%
_{1}}^{\dagger }d_{x,\uparrow ,\mathbf{R}_{2}}^{\dagger }+\mathrm{H.c.}\right] \text{,%
}  \label{eq:H1}
\end{equation}%
where $\mathbf{R}_{1}=\left( 0,0\right) $ and $\mathbf{R}_{2}=\left(
a,a\right) $. This term can be thought of as a proximity coupling to a bulk
superconductor. Note that since the pairing term is applied only to the $%
d_{xz}$ orbital, it couples to any (singlet) superconducting order parameter.

Fig. \ref{fig:4}a shows the induced bond pair amplitudes $\phi _{\mathbf{R},%
\mathbf{R}^{\prime }}=\langle d_{\alpha ,\uparrow
,\mathbf{R}}^{\dagger }d_{\alpha ,\downarrow ,\mathbf{R}^{\prime
}}^{\dagger }-d_{\alpha ,\downarrow ,\mathbf{R}}^{\dagger
}d_{\alpha ,\uparrow ,\mathbf{R}^{\prime }}^{\dagger
}+\mathrm{H.c.}\rangle $ for a $6\sqrt{2}a\times \sqrt{2}a$ (24
site) system with $\Delta =0.5$ and $U=4$. With the pair field
term (\ref{eq:H1}), the number of electrons in the system is not
conserved. Here, we
choose the chemical potential $\mu$ such that the
average number of electrons in the system is close to $42$, which
corresponds to a hole doping of $n=0.25$ per Fe site.
The pair amplitudes decay slowly with the distance 
from the edge. Near the middle of the system, the pair structure
is
shown in Fig. \ref{fig:4}b. One can see that under reflection about the $%
\left( 1,\pm 1\right) $ directions, the pair wave function is
\emph{even}, indicating A$_{1g}$--like pairing. The A$_{1g}$
structure appears already in smaller systems (down to
$2\sqrt{2}a\times \sqrt{2}a$).
The same pairing occurs in 
electron-doped systems and even in the undoped system, 
although here the pairing response is considerably weaker.
For $U=8$, the pairing symmetry is still A$_{1g}$, although the pairing
amplitude decays considerably faster than for $U=4$.

\begin{figure}[t]
\includegraphics[width=0.45\textwidth]{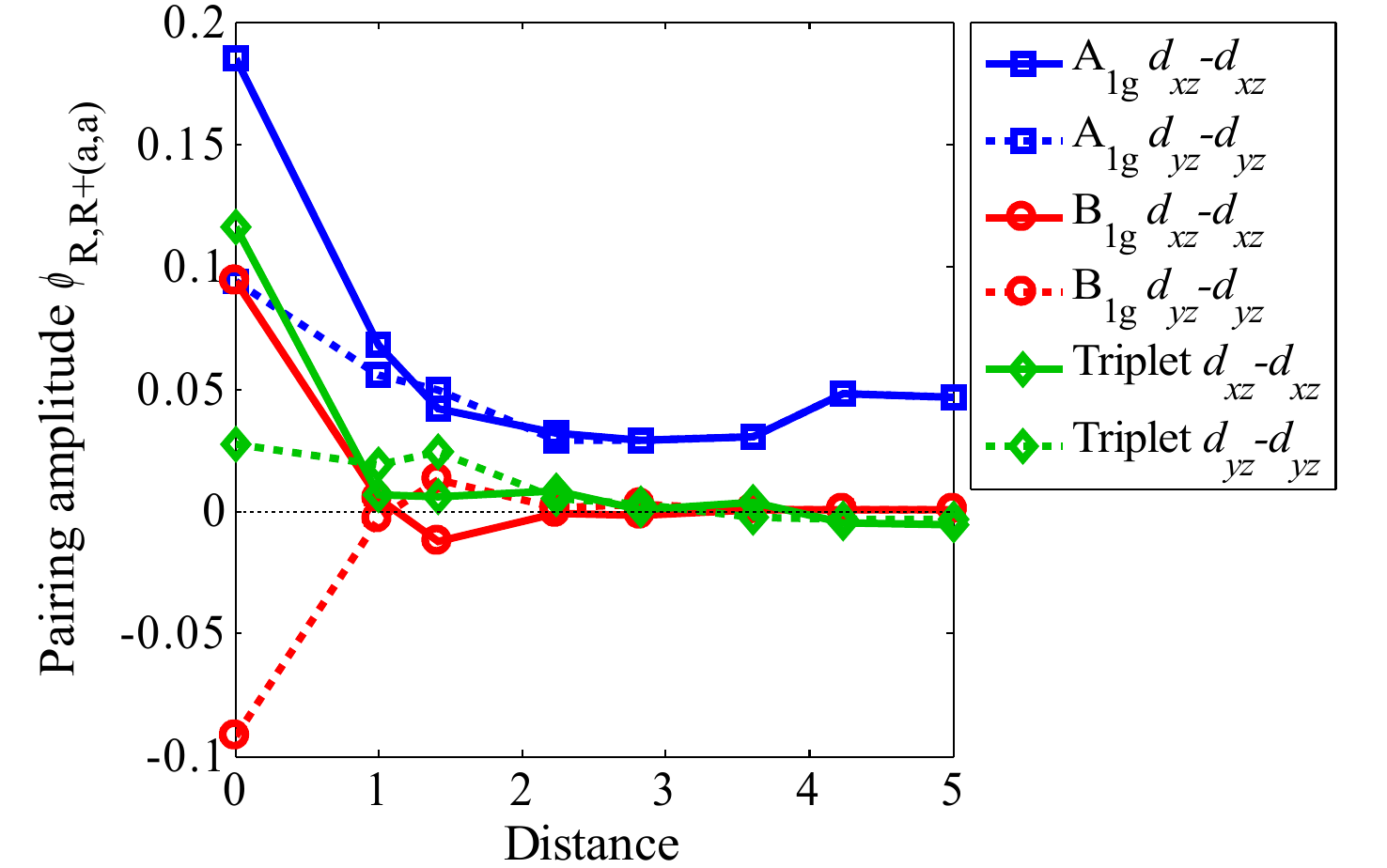}
\caption{(Color online.) Induced pair amplitudes on diagonal bonds in three $%
4\protect\sqrt{2}a\times \protect\sqrt{2}a$ calculations, in which A$_{1,g}$
($\square$), B$_{1,g}$ ($\circ$) and triplet ($\diamond$) pair fields are
applied to the edge (see text). Solid (dashed) lines correspond to $%
d_{xz}-d_{xz}$ ($d_{yz}-d_{yz}$) pairing, respectively.}
\label{fig:5}
\end{figure}

Fig. \ref{fig:5} shows the diagonal $d_{xz}-d_{xz}$ and $d_{yz}-d_{yz}$
pairing amplitude as a function of position, in calculations with various
pairing potentials applied at the edge, such that they excite different
pairing symmetries selectively. For A$_{1g}$, Eq. (\ref{eq:H1}) was used.
For B$_{1g}$, we have added an additional term for the $d_{yz}$ orbital,
with an opposite sign. This term does not couple to the A$_{1g}$ pairing
symmetry at all, because it is odd under reflection about $\left(
1,-1\right) $. In addition, we have 
applied a triplet pairing term [in which the sign in the square brackets of
Eq. (\ref{eq:H1}) is reversed]. Strictly speaking, the B$_{2g}$ ($D_{xy}-$%
like) symmetry is not distinct from A$_{1g}$. However, one can still think
about a \textquotedblleft B$_{2g}-$like\textquotedblright\ pair structure in
which the sign of the pair amplitude on $\left( 1,1\right) -$ and $%
\left( 1,-1\right) -$oriented bonds is opposite. To couple mostly
to B$_{2g}$, we have added to Eq. (\ref{eq:H1}) a bond pairing
term on the $\left( 0,0\right) -\left( a,-a\right)$ bond with an
equal magnitude and opposite sign. Alternatively, we have applied
a singlet $d_{xz}-d_{yz}$ pair field on the $(0,0)-(a,0)$ bond,
which also favors B$_{2g}$-like pairing. Comparing the response of
the different pairing symmetries, the A$_{1g}$ pairing clearly
decays much more slowly than the triplet and the B$_{1g}$
pairings. In the B$_{2g}$ calculations (not shown), the induced
pairing structure is B$_{2g}-$like very close to the edge, but
changes its character to A$_{1g}-$like further away into the bulk.

\emph{Discussion}-- In conclusion, the two--orbital diagonal
ladder model shows several robust features, which appear already
at very small system sizes. In electron and hole doped systems,
there is a clear tendency to form A$_{1g}$ ($S-$wave)
superconducting order. All other forms of order are
very weak. 
For the undoped system, a strong tendency toward antiferromagnetic ordering
is observed. However, two forms of magnetic order are in close competition
with each other: unidirectional \textquotedblleft stripe\textquotedblright\
order, of the sort found in experiment, and non-collinear \textquotedblleft
spin-checkerboard\textquotedblright\ (OM) order, which was found also in
mean-field calculations\cite{Lorenzana-short}. In our model, the
spin-striped state is stabilized for $U\gtrsim 18$, and the OM is stabilized
for smaller values of $U$.
The near degeneracy of these two states implies that 
small terms can tilt the balance one way or the other, which
raises the possibility that the OM state may be stabilized in some
member of the Fe-pnictide or chalcogen families. Finally, in the
undoped system there is a strong tendency towards nematic order.
This order is associated with the expectation values of bond
operators (\emph{e.g.} the local kinetic energy and spin-spin
correlations), rather than the difference of onsite
orbital occupations which were found to be small. 

\emph{Acknowledgement--} We are grateful to E. Dagotto, J.
Lorenzana, A. Moreo, F. Pollmann and W-F. Tsai for their comments
on this manuscript. DJS acknowledges the Center for Nanophase
Materials Science at ORNL, which is sponsored by the Division of
Scientific User Facilities, U.S. DOE, and thanks SITP at Stanford
for their hospitality. SAK was supported, in part, by the NSF
under grant DMR-0758356. EB was supported by the NSF under grants
DMR-0705472 and DMR-0757145. We acknowledge the Center for
Nanoscale Systems (CNS) (supported by NSF award no. ECS-0335765)
at Harvard University, for allocation of computer time through the
Odyssey cluster supported by the FAS Research Computing Group.
This work was supported in part by the NSF under the grant
PHY05-51164 at the KITP.

\bibliography{FeAs}

\end{document}